\documentclass[aps,prl,twocolumn,preprintnumbers,superscriptaddress,10pt]{revtex4}

\usepackage[dvips]{color}
\usepackage[normalem]{ulem}

 \usepackage{epsfig}

\begin{document}

\def\ov{\over}
\def\le{\left}
\def\ri{\right}
\def\ha{{1\over 2}}
\def\lam{{\lambda}}
\def\Lam{{\Lambda}}
\def\al{{\alpha}}
\def\ket#1{|#1\rangle}
\def\bra#1{\langle#1|}
\def\vev#1{\langle#1\rangle}
\def\det{{\rm det}}
\def\tr{{\rm tr}}
\def\Tr{{\rm Tr}}
\def\NN{{\cal N}}
\def\th{{\theta}}

\def\Om{{\Omega}}
\def \th{{\theta}}

\def \lam {\lambda}
\def \om {\omega}
\def \ra {\rightarrow}
\def \ga {\gamma}
\def\sig{{\sigma}}
\def\ep{{\epsilon}}
\def\apr{{\alpha'}}
\newcommand{\p}{\partial}
\def\LL{{\cal L}}
\def\TT{{\cal T}}
\def\tir{{\tilde r}}

\newcommand{\be}{\begin{equation}}
\newcommand{\ee}{\end{equation}}
\newcommand{\bea}{\begin{eqnarray}}
\newcommand{\eea}{\end{eqnarray}}
\newcommand{\nn}{\nonumber\\}

\title{An AdS/CFT calculation of screening in a hot wind}

\author{Hong Liu}
\affiliation{Center for Theoretical Physics, Massachusetts Institute
of Technology, Cambridge, MA 02139, USA}
\author{ Krishna Rajagopal}
\affiliation{Center for Theoretical Physics, Massachusetts Institute
of Technology, Cambridge, MA 02139, USA}
\affiliation{Nuclear Science Division, MS 70R319,
Lawrence Berkeley National Laboratory, Berkeley, CA 94720, USA}
\author{Urs Achim Wiedemann}
\affiliation{Department of Physics, CERN, Theory Division, CH-1211 Geneva 23, Switzerland}

\preprint{MIT-CTP-3757, CERN-PH-TH/2006-121}


\begin{abstract}
One of the challenges in relating experimental measurements of the suppression
in the number of $J/\psi$ mesons produced in heavy ion collisions to lattice QCD
calculations is that whereas the lattice calculations treat $J/\psi$ mesons at rest,
in a heavy ion collision 
a $c\bar c$ pair can have a significant velocity with respect to the hot fluid 
produced in the collision.  The putative $J/\psi$ finds itself in a hot wind.
We 
present the first rigorous non-perturbative calculation of the consequences of a wind 
velocity $v$ on the screening length $L_s$ for a heavy quark-antiquark pair
in hot ${\cal N}=4$ supersymmetric QCD. We find $L_s(v,T) = f(v)[1-v^2]^{1/4}/\pi T$ with
$f(v)$ only mildly dependent on $v$ and the wind direction. This
$L_s(v,T)\sim L_s(0,T)/\sqrt{\gamma}$ velocity scaling, if realized in QCD,
provides a significant additional source of
$J/\Psi$ suppression at transverse momenta which are high but within
experimental reach.
\end{abstract}
\maketitle


Twenty years ago, Matsui and Satz suggested that because the attraction between
a quark and an antiquark is screened in a deconfined quark-gluon plasma, the production
of $J/\psi$ mesons should be suppressed in sufficiently energetic nucleus-nucleus
collisions relative to that in
proton-proton or proton-nucleus collisions, since the screened interaction
between a $c$ and $\bar c$
immersed in a quark-gluon plasma would not bind them~\cite{Matsui:1986dk}.
In the intervening years, marked progress on many fronts has not changed this
basic qualitative picture. On the experimental side, we now have
data from the NA50 and NA60 experiments at the CERN SPS
and from the PHENIX experiment at RHIC that demonstrate the
existence of a suppression~\cite{ExperimentalRefs}. On the theoretical side, we now have ab initio calculations
of the temperature-dependent
potential between a color singlet heavy quark and antiquark
separated by a distance $L$~\cite{Kaczmarek:2004gv,Kaczmarek:2005ui}.
This potential is as at $T=0$ for small $L$, but begins to weaken for
$L$ larger than some $L_s$ and flattens at larger $L$.
The potentials obtained in these lattice calculations can be crudely characterized as
indicating $L_s \sim 0.5/T$ in hot QCD with two flavors of light quarks~\cite{Kaczmarek:2005ui}
and
$L_s\sim 0.7/T$ in hot QCD with no dynamical quarks~\cite{Kaczmarek:2004gv}.
Furthermore, lattice QCD calculations of the Minkowski space $J/\psi$ spectral
function itself have now been done in quenched QCD~\cite{Asakawa:2003re}, and early results in QCD
with dynamical quarks have also been reported~\cite{Morrin:2005zq}. These studies indicate that the
$J/\psi$ meson ceases to exist as a bound state above a temperature somewhere
between $1.5\,T_c$ and $2.5\,T_c$, in agreement with conclusions drawn based upon
the screening potential between static quarks~\cite{Karsch:2005nk}.

The multifaceted challenge, now, is
to make quantitative
contact between the lattice calculations and data from heavy ion collisions.
One significant difficulty is that the lattice calculations
treat a quark-antiquark pair in the quark-gluon plasma rest frame, whereas in a heavy
ion collision a $c\bar c$ pair is not produced at rest.  This challenge becomes more
acute in higher energy collisions: at LHC energies, $c\bar c$
pairs which if produced in vacuum would yield $J/\psi$ mesons with
transverse momenta many times their rest mass will be copious.  Even
in collisions at SPS and RHIC energies,  the collective flow developed by the hot
medium in which the $c \bar c$ pair finds itself is considerable.
A rigorous
determination of the $v$-dependence of the screening length $L_s(T)$ for
a heavy quark antiquark pair in a quark-gluon plasma moving with velocity $v$
would therefore be a significant advance.  We provide one, albeit for hot ${\cal N}=4$
super Yang-Mills theory.

${\cal N}=4$ super Yang-Mills (SYM) theory is a conformally invariant theory with two
parameters: the rank of the gauge group $N_c$ and the 't Hooft coupling
$\lambda = g_{\rm YM}^2 N_c$.  We shall define the screening length $L_s$
below, based upon an analysis of a fundamental Wilson loop
describing the dynamics of a
color-singlet quark-antiquark ``dipole'' moving with velocity $v$
along, say, the $x_3$-direction through the hot strongly interacting ${\cal N}=4$ SYM
plasma. In the
rest frame of the dipole, which sees a hot wind blowing in the $x_3$-direction,
the contour
${\cal C}$ of the Wilson loop is given by a rectangle with large
extension ${\cal T}$ in the $t$-direction, and short sides of
length $L$ along some spatial direction.
Evaluating
this Wilson loop (whether ultimately in QCD or in ${\cal N}=4$ SYM as we do here)
will teach us about the $L$-dependent color singlet quark-antiquark
potential and
hence allow us to define a screening length in the presence of a hot wind.

According to the AdS/CFT correspondence~\cite{AdS/CFT},
in the large-$N_c$ and large-$\lambda$ limits
the thermal expectation value
$\langle{W^F({\cal C})}\rangle$
for the Wilson loop in the absence of a wind velocity
can be calculated using the metric for a 5-dimensional curved
space-time describing a black hole in
anti-deSitter (AdS) space~\cite{Rey:1998ik}.  Calling the fifth dimension $r$,
the black hole horizon is at some  $r=r_0$
and we add a probe D3-brane extended along the $x_1$, $x_2$, $x_3$ directions
at some $r=\Lambda\gg r_0$. The external quarks described by the
Wilson loop are open strings ending on the probe brane.
The prescription for evaluating $\langle{W^F({\cal C})}\rangle$
is that we must find the extremal action surface in the five-dimensional AdS spacetime
whose boundary at $r=\Lambda$ is the contour
${\cal C}$ in Minkowski space $R^{3,1}$.  $\langle{W^F({\cal C})}\rangle$ is
then given by
$\exp[ i S ({\cal C})]=\exp[iE({\cal C}){\cal T}]$,
with $S$ the action of the extremal surface~\cite{Rey:1998ik}. For the time-like
Wilson loop we analyze, $S({\cal C})$  is proportional to the time ${\cal T}$, meaning that
$E({\cal C})$ can be interpreted as the energy of
the dipole.
In
the limit of infinitely heavy quarks, i.e. $\Lambda\rightarrow\infty$,  $E({\cal C})\propto \Lambda$ but this  divergence
comes from the $L$-independent self-energy of the quark or antiquark
taken separately.
We are only interested in the $L$-dependent part of $E({\cal C})$, which is finite in
the $\Lambda\rightarrow \infty$ limit, and we take
this limit henceforth~\footnote{At this point the present calculation becomes qualitatively
distinct (even in the $v\rightarrow 1$ limit) from the calculation of the light-like
Wilson loop 
used to determine the ``jet quenching parameter'' $\hat q$ in 
Refs.~\cite{Liu:2006ug,Buchel:2006bv}.
Although we can apply our calculation of screening in a hot wind
in the $v\rightarrow 1$ 
limit, it does {\it not} reduce to the calculation of $\hat q$ in this limit.
In this paper, we calculate the screening potential between infinitely
massive (mass $\propto \Lambda$, where 
we have taken $\Lambda\rightarrow\infty$)
test quarks. To reproduce the calculation of $\hat q$, we must instead
{\it first} take the $v \rightarrow 1$ limit 
at finite $\Lambda$, and only then are free
to take $\Lambda\rightarrow \infty$. The two limits do not commute. 
The string world sheet which in our
screening calculation is time-like becomes 
space-like for $r_0^4 \gamma^2 >\Lambda^4$,
meaning for high enough wind velocity.
Whereas a time-like world sheet
(and $\langle W \rangle \sim \exp[i S]$) has a sensible physical interpretation
as the calculation of an interaction energy and hence yields information
about screening, a space-like world 
sheet (and $\langle W \rangle \sim \exp[-S]$ with
$S$ real) has a sensible physical interpretation in the context of high
energy scattering in an eikonal approximation~\cite{Liu:2006ug}.
Finally, it is possible 
to show that upon
starting with $\eta$ and $\Lambda$ finite and 
taking the $\Lambda\rightarrow \infty$ limit while 
keeping $r_0^4\gamma^2 \gg \Lambda^4$,
one does obtain the jet quenching parameter $\hat q$ of Ref.~\cite{Liu:2006ug}.
In calculating $\hat q$ 
using this limiting procedure, there is only a single extremal world sheet: the
``trivial world sheet'', found and discarded in Ref.~\cite{Liu:2006ug}, 
does not even arise.}.

To describe a hot wind in the $x_3$-direction, we boost the five-dimensional
AdS black hole metric, obtaining
\be
ds^2 = - A dt^2 + 2B dt dx_3 + C dx_3^2 + {r^2} (dx_1 + dx_2^2)+
f^{-1}dr^2
\label{windmetric}
 \ee
where $f = {r^2 \over R^2} \left(1 - {r_0^4 \over r^4} \right)$ with $R$ the curvature radius
of the AdS space and where we have defined
\be
A = {r^2 \ov R^2}  - {r_1^4 \ov r^2 R^2}, \quad B = {r_1^2 r_2^2 \ov r^2 R^2},\quad
C = {r^2 \ov R^2}  + {r_2^4 \ov r^2R^2}
\label{ABCdefn}
 \ee
and
$r_1^4 = r_0^4 \cosh^2 \eta = r_0^4 \gamma^2$ and  $r_2^4 = r_0^4 \sinh^2 \eta =
r_0^4 \gamma^2 v^2$,
with $v$ the velocity of the wind, $\eta$ its rapidity, and $\gamma=1/\sqrt{1-v^2}$.
Here,
the temperature $T$ of the Yang-Mills theory at $v=0$
is given by the Hawking temperature of the
black hole,  $T = \frac{r_0}{ \pi R^2}$,
and $R$ and the string tension
$1/2\pi\alpha'$ are related to the t'Hooft coupling by $\frac{R^2}{\alpha'} = \sqrt{ \lambda}$.

The short side of ${\cal C}$ can be chosen to lie in the $(x_1,x_3)$ plane, at an
angle $\theta \in [0,\pi/2]$ relative to the $x_3$-direction. $\theta$ is the angle
between the dipole and the wind direction in the dipole rest frame.
We parameterize the surface whose action $S({\cal C})$ is to be
extremized by  $x^\mu = x^\mu (\tau,\sigma)$,
where $\sigma^\alpha = (\tau, \sigma)$ denote world sheet coordinates.
The Nambu-Goto action for the string world sheet is given by
 \begin{equation}
S ={1 \over 2 \pi \alpha'} \int d\sigma d \tau \, \sqrt{ - \det g_{\alpha
\beta}}
\label{NGaction}
 \end{equation}
with $g_{\alpha \beta} = G_{\mu \nu} \partial_\alpha x^\mu \partial_\beta x^\nu$
the induced metric on the world sheet. This action is invariant under coordinate
changes of $\sigma^\alpha$, and we can set $\tau = t$ and
$\sigma = x_1$~\footnote{With the exception of a wind parallel
to the dipole ($\theta=0$), which can be formulated easily as a special case.}.
Note that
$\sig$ takes values $-{L \ov 2} \sin \th < \sig < {L \ov 2} \sin\th$.
Since ${\cal T} \gg L$, we can assume that the surface is
translationally invariant along the $\tau$ direction, i.e.
$x^\mu (\sigma, \tau)=x^\mu (\sigma)$.
The Wilson loop lies at constant $x_2$, so $x_2(\sigma)={\rm const}$.
The boundary condition for $x_3(\sigma)$ can be taken as
$x_3(\pm\frac{L}{2} \sin \theta)= \pm {L \ov 2} \cos \theta$. For the bulk
coordinate $r(\sigma)$, we implement the requirement that the world sheet
has ${\cal C}$ as its boundary by imposing
$r \left(\pm  {L \over2}  \sin \theta \right) = \Lam \to \infty$. It proves convenient
to define $y\equiv r/r_0$ and $z\equiv x_3 r_0/R^2 = x_3\pi T $
and hence $\ell \equiv  Lr_0/R^2=L\pi T$ and to define a rescaled
$\tilde\sigma= \sigma r_0/R^2$ and promptly drop the tilde.
The
boundary conditions become
\be
y\left(\pm \frac{\ell}{2} \sin \th \right)=\infty \qquad
z\left(\pm \frac{\ell}{2} \sin \th\right)=\pm{\ell \ov 2} \cos \th \ .
\label{boundaryconditions}
\ee

\begin{figure}[t]
\vskip-0.15in
\hfill\hskip-0.6in \includegraphics[scale=0.2,angle=-90]{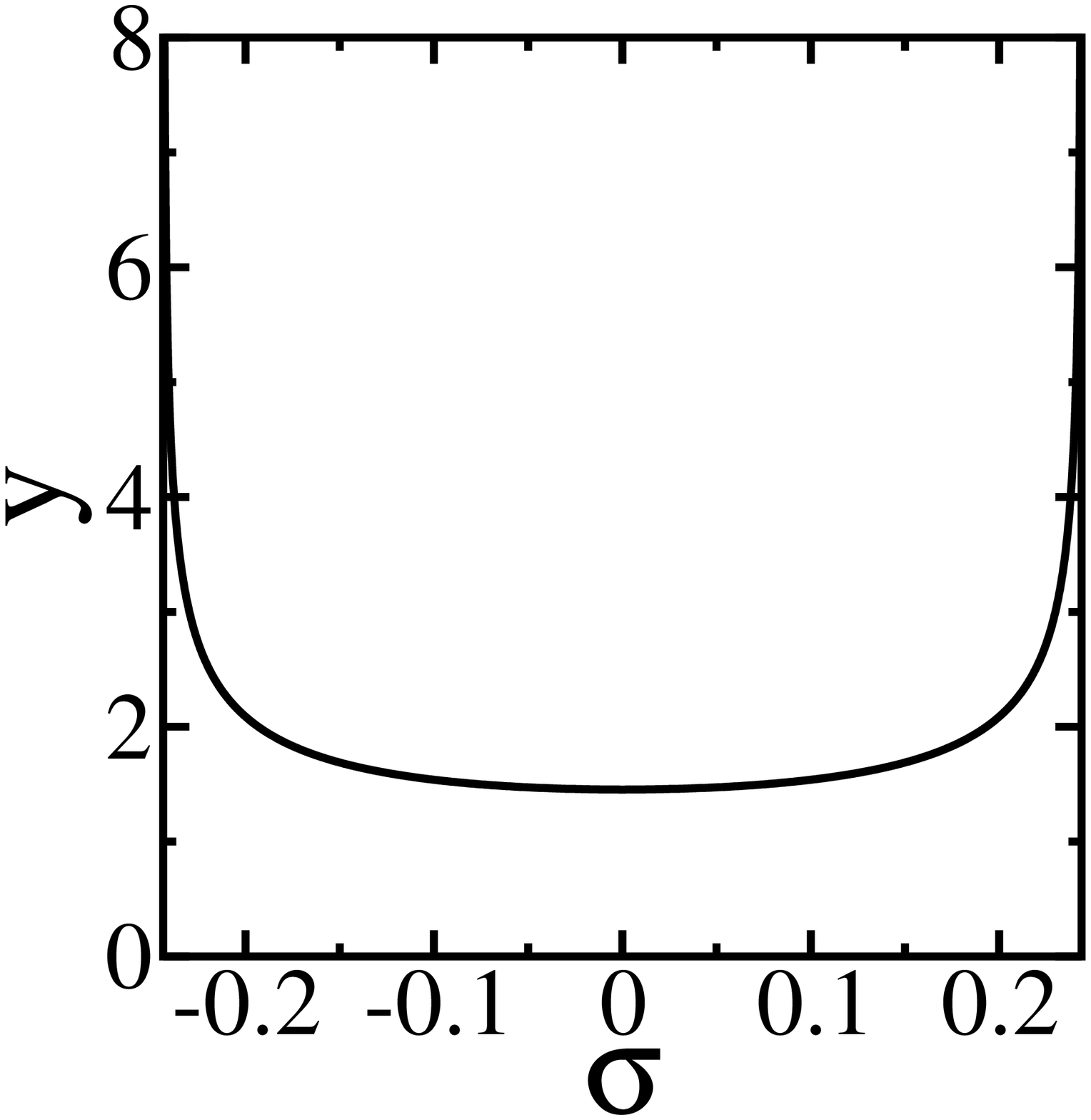}\hskip-0.4in \includegraphics[scale=0.2,angle=-90]{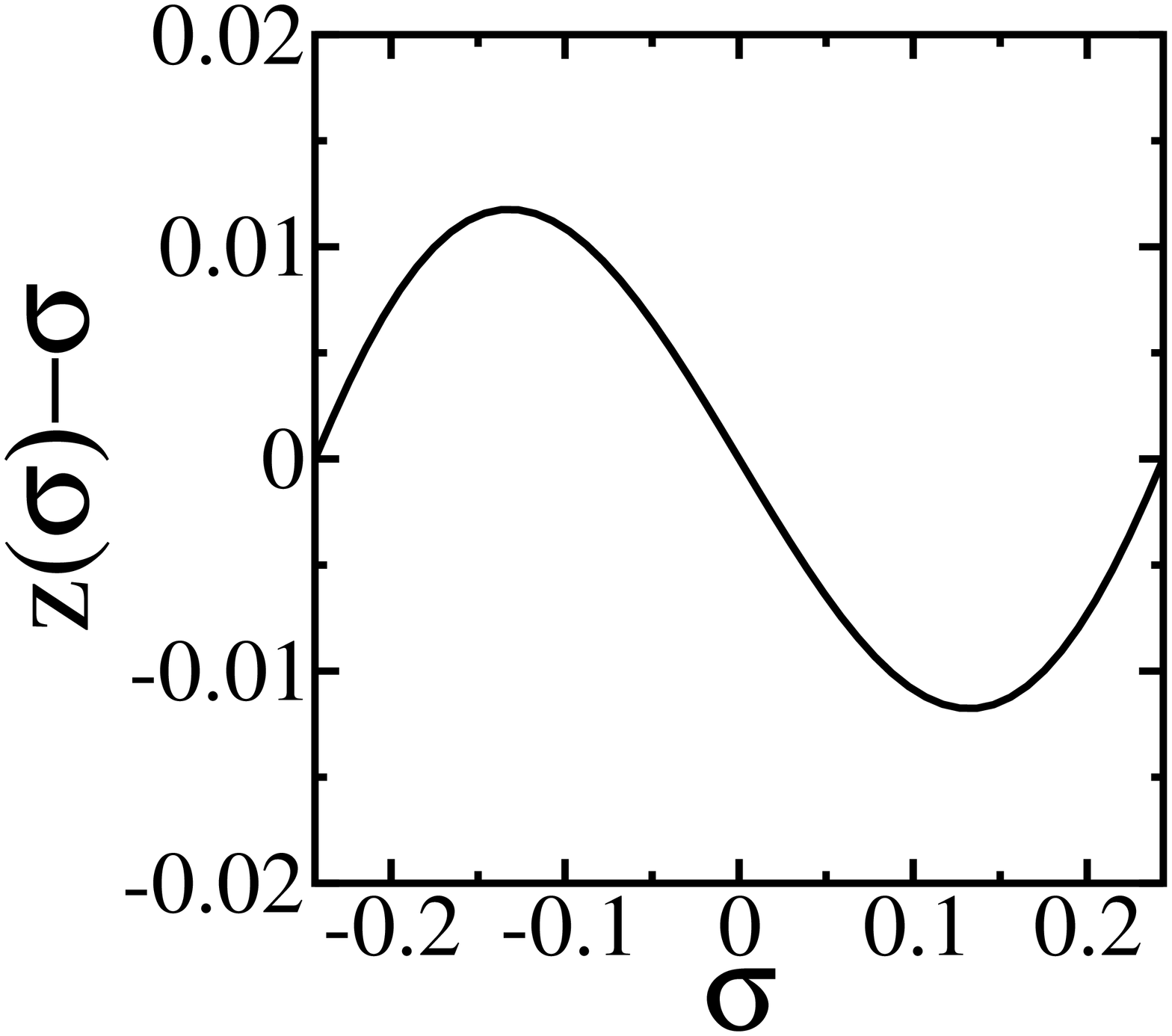}\hskip-0.1in\hfill
\caption{String world sheet for wind with velocity $v=0.7$ blowing at an
angle $\theta=45^\circ$ relative to the dipole.  The solution has integration
constants $p=1.325$ and $q=1.109$, which correspond to $\theta=45^\circ$
and $\ell=0.689$. (This $\ell$ is the maximum possible
for this $v$ and $\theta$.)  $\sigma\propto x_1$
extends from $-(\ell/2)\sin\theta$ to $+(\ell/2)\sin\theta$.
(a) $y(\sigma)$. (b) $z(\sigma)-\sigma$ is the deviation of the string world sheet
away from $z=\sigma$, the straight line at $\theta=45^\circ$
between the quark and the antiquark.
}
\vskip-0.05in
\end{figure}

The action (\ref{NGaction}) now takes the form
 \be \label{binact}
\frac{S}{\cal T} = E = K  \int_0^{l \ov 2} d \sig
\; {\cal L}
 \ee
with the constant prefactor given by
$K = \sqrt{ \lambda }T $
and with the Lagrangian
\be \label{Lagrangian}
 {\cal L} = \sqrt{\le(y^4 - {\cosh^2 \eta } \ri) \le(1 + {y'^2 \ov y^4
 - 1} \ri) + z'^2\le(y^4-1\ri)}\ ,
 \ee
where henceforth $'\equiv \frac{\partial}{\partial\sigma}$.
The equations of motion obtained from this Lagrangian can be cast in the form
\bea
q^2 y'^2 &=& (y^4-\cosh^2\eta)(y^4-1-p^2) - q^2(y^4-1)\ ,\qquad\label{yeqn}\\
q^2 z'^2 &=& p^2\left(\frac{y^4-\cosh^2\eta}{y^4-1}\right)^2\ ,\label{zeqn}
\eea
where $p$ and $q$ are integration constants.
Since $y'$ depends only on $y$
and the boundary condition for $y$ is
symmetric under $\sigma\leftrightarrow-\sigma$, we find
$y({\sigma})$ to be an even function of $\sigma$:   it descends ($y'<0$)
for $-\ell/2<\sigma<0$, turns around at $\sigma=0$ where $y'=0$, and then ascends.
Since $z'$ depends
only on $y$ and the boundary condition for $z$ is antisymmetric in
$\sig \to -\sig$, we find $z (\sig)$ to be an odd function
of $\sig$. In particular, $z =0$ at $\sig =0$.

Eqs. (\ref{yeqn}) and (\ref{zeqn}) with
the boundary conditions (\ref{boundaryconditions}) can be
integrated numerically and in Fig.~1 we present an example.
Somewhat counterintuitively, the projection of the string
world sheet onto the $(x_1,x_3)$-plane is not a straight line connecting
the quark and the antiquark, but rather has a sinusoidal form.
This
behavior arises for all values of $\theta$ except $\theta=0$ or $\pi/2$
(wind parallel or perpendicular to the dipole) for which the projection
is indeed a straight line.   Furthermore, we see that even though there is a wind
blowing in the $z$-direction,
$y(\sigma)$ is even and $z(\sigma)$ is odd for any angle $\theta$, meaning that
the string world sheet is not dragged at all by this wind.
This conclusion is
antithetical to that for the world sheet of an isolated string ending on a single quark
or antiquark, analyzed in Ref.~\cite{Herzog:2006gh}, and in qualitative agreement with the conclusion
that ``mesons feel no drag'' reached in a different context in Ref.~\cite{Peeters:2006iu}.
The $\theta=\pi/2$ case
is particularly simple
to analyze because the integration constant $p$ vanishes~\footnote{If $p$ and hence $z'$ were nonzero, the 
boundary condition
$z(+\ell/2)=z(-\ell/2)$ would then require that there be point(s) at which $z'=0$.  From
(\ref{yeqn}) and (\ref{zeqn}) we see that at such a point $y'^2$ would be negative.
Hence, $p=0$.}.
With $p=0$, $z(\sigma)$ is constant, starkly making 
the point that the world sheet is not
dragged by the wind blowing in the $z$-direction.

We shall find that the screening length behaves similarly for all values
of $\theta$.  For purposes of illustration, we present the analysis for
$\theta=\pi/2$. With $z(\sigma)=$ constant, we need only
analyze the shape of the world sheet in the $r$-direction, governed
by (\ref{yeqn}) which we can write as
$y' = {1 \ov q} \sqrt{(y^4- 1) (y^4 - y_c^4)}$
 upon defining
$y_c^4 = \cosh^2 \eta + q^2$.
The world sheet stretches from $y=\infty$ down to $y=y_c$,
with $y' (0) =0$ and $y(0)=y_c$.
The integration constant $q$ can now be determined from the equation ${\ell \ov 2} =
\int^{\ell \ov 2}_0 d \sig$, which 
becomes
 \be \label{ellvsq}
\ell  =
 {2 q} \int_{y_c}^{\infty} dy \, {1 \ov
\sqrt{(y^4- y_c^4) (y^4 - 1)} }\ .
 \ee
The energy can be written as
 \be \label{actionresult}
E =K \int_{y_c}^\infty dy \; { y^4 - \cosh^2\eta \ov \sqrt{(y^4
-1)(y^4-y_c^4)}} \ ,
 \ee
and can be made finite by subtracting the self-energy
of an isolated quark and antiquark.

\begin{figure}[t]
\vskip-0.25in
\includegraphics[scale=0.23,angle=-90]{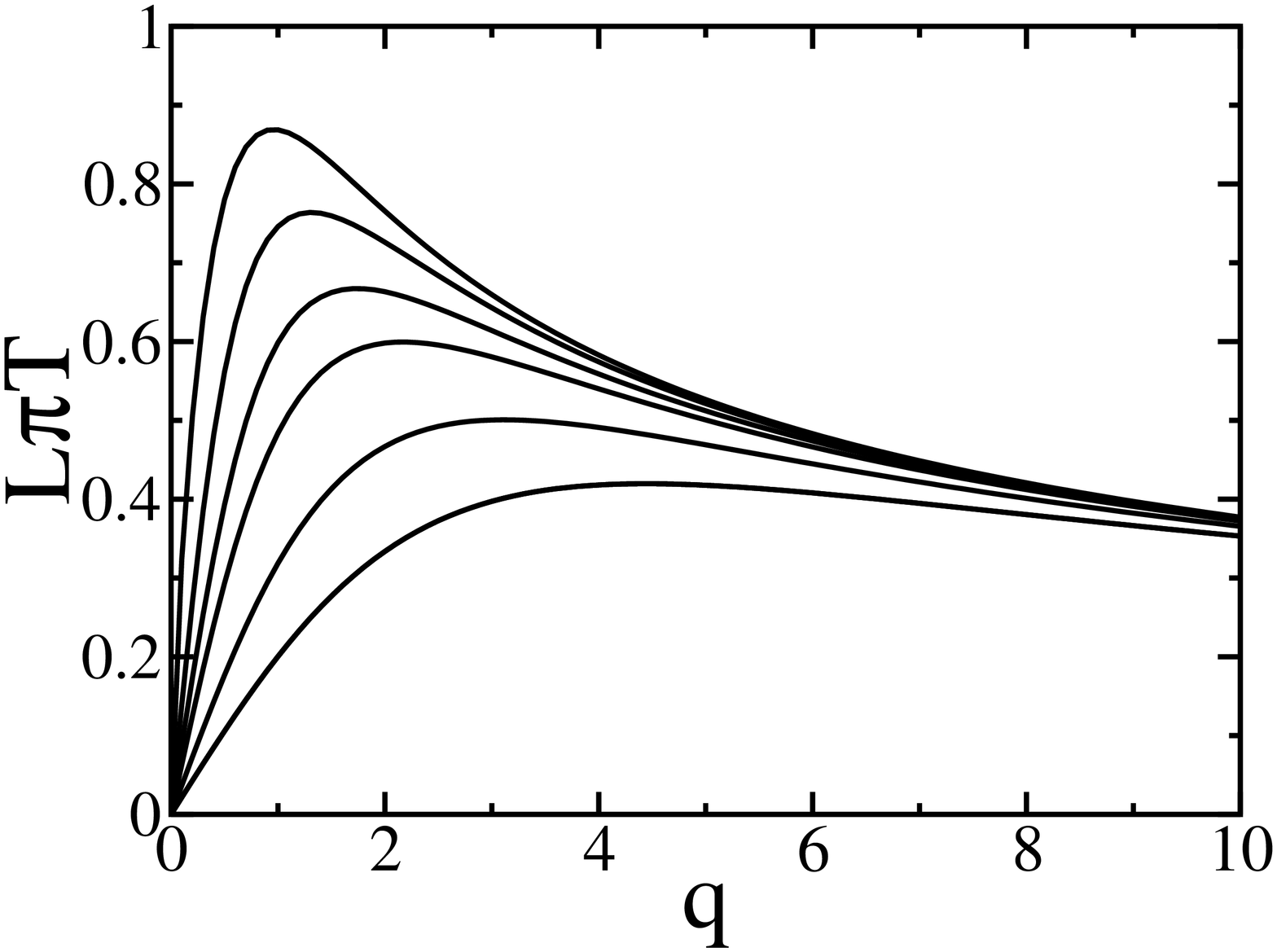}
\vskip-0.2in
\includegraphics[scale=0.23,angle=-90]{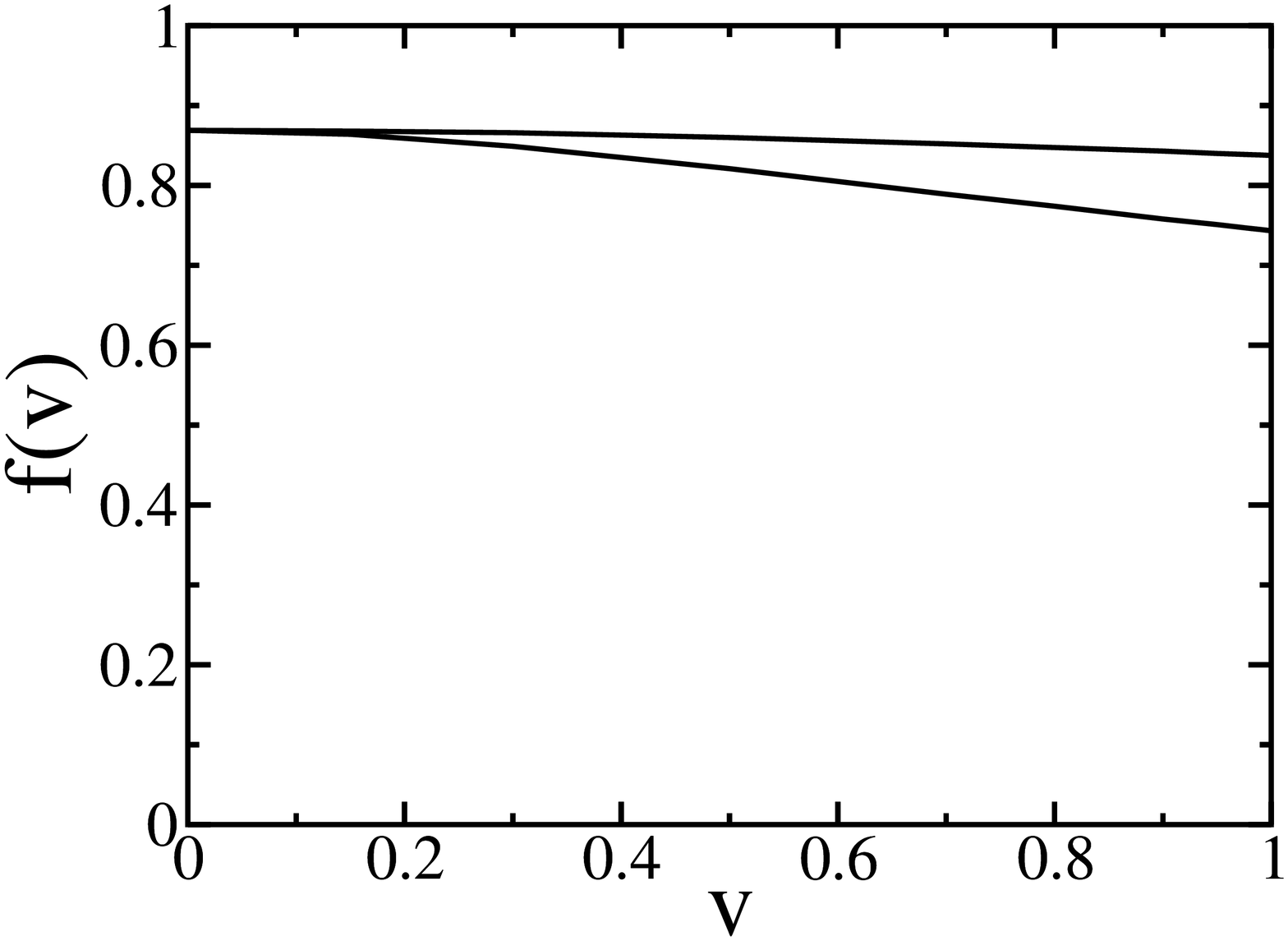}
\vskip-0.1in
\caption{The dependence of the screening length on the wind velocity $v$. (a)~$\ell=L\pi T$, given
in Eq.~\ref{ellvsq}, as a function of the integration constant $q$ for six different
values of $v$: 0, 0.5, 0.7, 0.8, 0.9, 0.95 (top to bottom).  We see the peak of this curve, $\ell_{\rm max}$,
dropping with $v$.  All curves are for a wind blowing in the direction
perpendicular to the dipole. (b)~$f(v)=\ell_{\rm max}\sqrt{\gamma}=L_s \pi T \sqrt{\gamma}$
versus $v$ for a wind blowing perpendicular to the dipole (lower curve) or
parallel to the dipole (upper curve).  
}
\vskip-0.05in
\end{figure}

In Fig.~2a we plot $\ell$ as a function of $q$ from Eq.~(\ref{ellvsq})
for several values of $\cosh^2\eta$.  Analysis of Eq.~(\ref{ellvsq}) shows
that, for any $\cosh^2 \eta$, $\ell$ decreases at large $q$ like $\ell\approx 1.198/\sqrt{q}$.
This corresponds to an
energy $E\propto - K \sqrt{q} \propto - K/\ell \propto -\sqrt{\lambda}/L$,
and thus describes
a quark and antiquark separated by a small distance $L$
interacting via an attractive Coulomb potential.  Small values of $L$ can also be
achieved by choosing $q\rightarrow 0$, as $\ell\propto q$ at small $q$.
These world sheets have $E\propto {\rm const}+L^2$,
meaning that these small $L$ solutions have much higher energy than
those describing Coulomb attraction, and so are not of interest.
Since $\ell$ is everywhere positive
and goes to zero for both small and large $q$, at some $q_m$ it has a maximum value
$\ell_{\rm max}$ as illustrated in Fig.~2a.  For $\ell>\ell_{\rm max}$,
no extremal world sheet bounded  by the Wilson loop ${\cal C}$ exists, with the exception
of disconnected world sheets ``hanging from'' the quark or the antiquark Wilson line separately,
describing their self energies.  Hence, for $\ell>\ell_{\rm max}$ there is {\it no} $L$-dependent
potential between the quark and antiquark~\footnote{In this theory as in QCD one
expects a residual attraction that falls exponentially with $L$ for  $L>L_s$.  Seeing
such effects (which are nonperturbative in $\alpha'$)
requires  analysis beyond extremizing the Nambu-Goto
action.}.
We can therefore define a screening length $L_s$
by $L_s \equiv \ell_{\rm max}/\pi T$~\footnote{Alternatively, 
we could define $L_s$ as the length $L_c$ below which
the nontrivial extremal world sheet has less energy than two disconnected world
sheets hanging from the quark and antiquark Wilson lines separately. 
$L_c$ is of order 10\%  smaller than our $L_s$ at $v=0$; they become equivalent
at high enough $v$.}.
Analysis of (\ref{ellvsq}) shows that for $v=0$, $q_m=0.96$ and $\ell_{\rm max}=0.869$
are numbers of order 1
while for large $\cosh^2\eta$, $q_m\approx 1.42 \cosh\eta $
and $\ell_{\rm max}\approx 0.743/\sqrt{\cosh\eta}=0.743(1-v^2)^{1/4}$.  This motivates
writing the screening length as
\be
L_s = \frac{f(v)}{\pi T} (1-v^2)^{\frac{1}{4}}\ ,
\label{fdefn}
\ee
in so doing defining the function $f(v)$.  We plot $f(v)$ in Fig.~2b, and find
that its velocity dependence is mild meaning that the dominant $v$-dependence of $L_s$
is the factor $(1-v^2)^{1/4}$.
For
general $\th$, the results are very similar: the interaction is
screened for $L>L_s$ where $L_s$ is given by (\ref{fdefn}) with a
slightly different $f(v)$, shown in Fig.~2b for a wind blowing
parallel to the dipole.  As the angle $\theta$ between wind and
dipole changes from $0^\circ$ to $90^\circ$, $f(v)$ interpolates
between the two curves in Fig.~2b.

Our central result is that,
in ${\cal N}=4$ SYM theory, the
dominant dependence of the screening length of a dipole in a hot wind on
the wind velocity is 
$L_s(v,T)\sim L_s(0,T)/\sqrt{\gamma}$, with the remaining weak
dependence described by the function $f(v)$ in Fig.~2b.  
The dominant velocity dependence suggests that
$L_s$ should be thought of
as $\propto$ (energy density)$^{-1/4}$, since the energy density increases like $\gamma^2$
as the wind  velocity is boosted.
It turns out that $L_s(0,T)$ is within a factor of  two of that for QCD.
Although it would be interesting to see whether $f(0)$ is closer to that in QCD in
more QCD-like theories with gravity duals, given the availability of reliable lattice
calculations of the screening potential in QCD itself this is not a pressing issue.
It would certainly be interesting to see whether
$L_s(v,T)\sim L_s(0,T)/\sqrt{\gamma}$ persists in other theories
with gravity duals, as this would support its
applicability to QCD~\footnote{In a hot wind, the large-spin
mesons in the confining, nonsupersymmetric theory studied by
Peeters, Sonnenschein and Zamaklar~\cite{Peeters:2006iu}
dissociate beyond a maximum wind velocity.  The relation between
the size $L$ of these mesons and their dissociation velocity $v$
is consistent with $L\propto (1-v^2)^{1/4}$, in qualitative
agreement with the result we have obtained analytically in a
simpler setting.}. 

\begin{figure}[t]
\includegraphics[scale=0.23,angle=0]{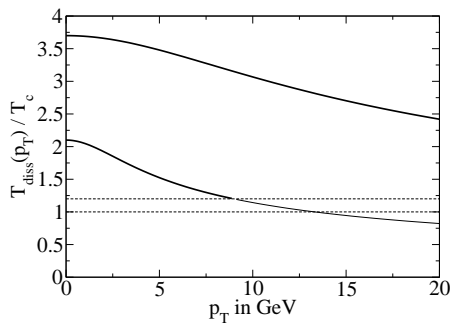}
\caption{A $1/\sqrt{\gamma}$-velocity scaling of the screening length
in QCD would imply a $J/\Psi$ dissociation temperature 
$T_{\rm diss}(p_T)$ that decreases significantly with $p_T$, while
that for the heavier $\Upsilon$ is affected less at a given $p_T$.  
The $\Upsilon$ curve is schematic:
we have increased $T_{\rm diss}(0)$ over that for the $J/\psi$ 
by a factor corresponding to its smaller size in vacuum.
}
\label{fig3}
\vskip-0.05in
\end{figure}

If the velocity-scaling of $L_s$ that we have discovered holds for
QCD, it will have qualitative consequences for
quarkonium suppression in heavy ion collisions.  For illustrative 
purposes, consider the explanation of the  $J/\Psi$ suppression
seen at SPS and RHIC energies proposed in Refs.~\cite{Karsch:2005nk,Satz}: lattice 
calculations
of the $q\bar q$-potential
indicate that the $J/\Psi$(1S) state dissociates at a temperature 
$\sim 2.1 T_c$ whereas the excited  $\chi_c$(2P) and
$\Psi'$(2S) states cannot survive above $\sim 1.2 T_c$; so, if
collisions at both
the SPS and RHIC reach temperatures above $1.2 T_c$ but not above
$2.1 T_c$, the experimental facts (comparable anomalous 
suppression of
$J/\Psi$ production at the SPS and RHIC) can be understood
as the complete loss  of the ``secondary'' $J/\Psi$'s that
would have arisen from the decays of the excited states, 
with no suppression 
at all of  $J/\Psi$'s that originate as $J/\Psi$'s.
Taking eq.~(\ref{fdefn}) at face value, the temperature  $T_{\rm diss}$ needed to 
dissociate the $J/\Psi$ decreases $\propto (1-v^2)^{1/4}$. As can be 
seen from Fig.~\ref{fig3}, this indicates that $J/\Psi$ suppression at RHIC will increase
markedly (as the $J/\Psi$(1S) mesons themselves dissociate)
for $J/\Psi$'s with transverse momentum $p_T$ above some threshold
that is at most $\sim 9$~GeV 
and would be $\sim 5$ GeV if the temperatures reached at RHIC are 
$\sim 1.5 T_c$~\footnote{Any $J/\Psi$ mesons
formed by recombination will have transverse momenta much lower
than those at which our calculation is relevant. Also, it is only at much
higher $p_T$ that one has to take into account the possibility that
$J/\Psi$ mesons could form outside the hot medium~\cite{Karsch:1987zw}.}.
These illustrative considerations point to a novel quarkonium suppression pattern
at transverse momenta above 5 GeV, a regime that is
within experimental reach of future high-luminosity runs at RHIC and that will be
studied thoroughly at the LHC.   If the temperatures reached at the LHC
are, say, $\sim 3 T_c$,
the LHC could discover $\Upsilon$ suppression, but only at high enough $p_T$.

\begin{acknowledgments}

HL is supported
in part by the A.~P.~Sloan Foundation and the U.S. Department
of Energy (DOE) OJI program.  
Research
supported in part by
the DOE
under
cooperative research agreement
\#DF-FC02-94ER40818.

\end{acknowledgments}



%

\end{document}